# Comparing and Contrasting Vibrational Wavepacket Dynamics and Impulsive Stimulating Raman Scattering Descriptions of Pump-Probe Spectroscopy: A Theoretical Study

Subho Mitra and Arijit K. De*

*Condensed Phase Dynamics Group*, Department of Chemical Sciences, Indian Institute of Science Education and Research Mohali, Knowledge City, Sector 81, SAS Nagar, Punjab, 140306

*Email: akde@iisermohali.ac.in

## Abstract

We simulate a third-order nonlinear signal in a pump-probe spectroscopy from the interference between first- and second-order wavepackets (WPs), as well as from a state-to-state transition for Stokes and coherent anti-Stokes pathways in the context of impulsive stimulated Raman scattering (ISRS) excitation. We present a detailed step-by-step description of both methods. The results obtained from these two methods are compared and contrasted through simulations of the excited-state absorption signal. While within the ISRS framework, vibrational dynamics is often attributed primarily to coherences between adjacent vibrational levels (i.e. $\Delta v = \pm 1$) in the excited electronic states, our results show that coherences involving non-adjacent vibrational levels (i.e. $\Delta v \neq \pm 1$) needs to be calculated for a better agreement with the WP approach. We also show that for the specific choice of pump/probe spectral bandwidths, the coherent anti-Stokes pathway majorly contributes to the observed signal.

## 1. Introduction

Light-induced functionalities arise from electronically excited states, whose time evolution governs photophysical and photochemical processes, such as energy and charge transfer [1-3]. Understanding excited-state dynamics in molecular systems is crucial for identifying the elementary events underlying these processes, such as internal conversion, intersystem crossing, and dissociation. Since these events occur on ultrafast timescales, real-time observation requires the use of time-resolved spectroscopy, for example, the widely used pump-probe spectroscopy, in which an ultrashort pump pulse initiates photoexcitation and a time-delayed probe pulse interrogates the changes induced by the pump pulse [4-9].

Owing to the large spectral bandwidth of an ultrashort pulse, many vibrational states are simultaneously excited. Apart from populating these vibrational states, an ultrashort laser pulse prepares a nonstationary superposition among them, i.e., a vibrational wavepacket (WP), on an excited-state potential energy surface (PES), whose subsequent propagation reflects the underlying topology, anharmonicity, and vibronic couplings. Thus, an understanding of the observed signal requires consideration of coupled electronic and nuclear motion, which naturally invokes a WP description following ultrashort laser excitation [10,11]. WP dynamics using pump-probe technique have been studied for decades as a powerful approach to track ultrafast nuclear and electronic motion in molecules [12-19]. Here, the vibrational coherences are generated from a single pump that interacts twice, and the probe pulse interacts once, resulting in the third-order ($\chi^{(3)}$ process) non-linear signal (therefore, it is known as 'four wave-mixing' process). In pump-probe geometry, the signal is heterodyne-detected with the probe [20]. It is important to note that, under impulsive excitation, depending on the wavelength of the pump pulse (i.e., resonant/non-resonant excitation process), the vibrational WPs are created both in the electronic ground and excited states. However, we restrict our discussion to the vibrational coherences created in the excited electronic state only.

With respect to the wavelength of the probe pulse, these vibrational coherences appear on the red and blue sides of the spectrum, following two distinct Liouville pathways, i.e., the Stokes and the coherent anti-Stokes pathways, as shown in **Figure 1** (for all possible combinations of an excited-state absorption (ESA) process). This description forms the basis of understanding impulsive stimulated Raman scattering (ISRS) [21-44]. This can be easily implemented in a pump/broadband probe setup, where multiple frequency components under the probe's flat spectral profile may contribute equally. Consequently, the red and blue parts of the spectrum can yield signals of comparable magnitude for the two pathways that emerge at the same detection wavelength. As these signals are anti-phased to each other, their temporal overlap leads to constructive or destructive interference (depending on the relative phase of the red and blue parts of the probe's spectrum), giving rise to single or multiple spectral nodes which has been a focus of recent studies [31-44].

On the other hand, while broadband ISRS has been widely employed to probe vibrational coherences in macromolecules, the resulting dynamics typically involve numerous coupled vibrational modes, leading to considerable complexity. It is therefore instructive to explore key aspects of this technique using simpler, well-defined model systems, where the underlying mechanisms can be more transparently understood. In this context, diatomic molecules such as molecular I₂, whose energetics and dynamics are extensively documented, serve as ideal test systems, particularly for a comparative study through numerical simulation.

Thus, from a fundamental point of view, it is intuitive to compare and contrast these two descriptions, i.e., vibrational WP dynamics and ISRS, in a diatomic molecule like I₂, which we explore in this article. We numerically simulate the pump-probe spectra in molecular iodine, considering only the ESA from the $B$ to the $E$ state. We focus on simulating the spectrally-resolved ISRS signal using a 'state-to-state excitation' approach that accounts for the two distinct pathways, and then compare the cumulative signal with that obtained from WP dynamics approach. Considering our specific choice of spectral bandwidths of pump and probe pulses, we show that the WP dynamics predominantly reflect signatures of a coherent anti-Stokes pathway.

## 2. Model and methods

### 2.1 Model

The simulation study is carried out by modelling the ground (X) and excited electronic states (B and E) of the diatomic iodine (I₂) system using Morse oscillator potential. The potential energy curves for the ground and excited states have the form,

$$V(r) = D_e \left[1 - e^{-\alpha(r - r_{eq})}\right]^2 \qquad (1)$$

where $D_e$ is the dissociation energy, and $\alpha$ is the well depth, $\omega_0 = \sqrt{\frac{2D_e \alpha^2}{m}}$ and $\chi_e = \frac{\hbar \omega_0}{4 D_e}$ refer to the respective vibrational mode and anharmonicity of the PES. For molecular I₂, the PESs parameters for the ground (X ($^1\Sigma_g^+$)) and first-excited state (B ($^3\Pi_{0u}^+$)) are taken from reference 45 [45] and for the second-excited state (E ($^3\Pi_{0g}^+$)), it is taken from reference 46 [46] and can be found in Table I. The B and E states are considered around 15,770 cm⁻¹ and 41,411 cm⁻¹ higher in energy than the X state [46]. The reduced mass (m) of I₂ is taken as 63.452 amu. The transition dipole moments $\mu_{BX}$ for both $B \leftarrow X$ transition and $\mu_{EB}$ for $E \leftarrow B$ transition is chosen as 0.461 D [47].

Table I. Parameters used in the simulation for molecular I₂

| References | Potential | Equilibrium bond length $r_{eq}$ (Å) | Dissociation energy $D_e$ (cm$^{-1}$) | $\alpha$ (Å$^{-1}$) |
|---|---|---|---|---|
| [45] | I - I (X) | 2.656 | 12550 | 1.871 |
| [45] | I - I (B) | 3.016 | 4503 | 1.850 |
| [46] | I - I (E) | 3.647 | 30808 | 0.560 |

The electric fields for both pump and probe, centered at $t = t_0$, have the general form as follows,

$$E(t) = \tfrac{1}{2}(E_0\, e^{-[(t-t_0)/T_0]^2}\, e^{-i\omega_0 t}\, e^{i\varphi} + \text{c.c.}) \quad (2)$$

where $E_0$, $\omega_0$ and $\varphi$ are the respective amplitude, frequency of oscillation, and the absolute phase of the field. $T_0$ is related to the full width half maxima (FWHM) of the intensity profile (i.e. $I(t) = |E(t)|^2$) of the field $\Delta t$ of the pulse as $T_0 = \frac{\Delta t}{\sqrt{2\ln 2}}$. The temporal intensity FWHM $\Delta t$ is related to spectral bandwidth $\Delta \omega$ through the time-bandwidth product,

$$\Delta t \Delta \omega = 4 ln 2 \quad (3)$$

where the spectral bandwidth is written as $S(\omega) = |\tilde{E}(\omega)|^2$ and $\tilde{E}(\omega)$ is the Fourier transform of $E(t)$. Now, for a transform-limited Gaussian pulse, the time-bandwidth product can also be alternatively written as,

$$\Delta t \Delta \nu = 0.441 \quad (4)$$

If the bandwidth is converted to wavenumber (cm$^{-1}$), the last equation modifies to,

$$\Delta t \Delta \tilde{\nu} = \frac{0.441}{c} \quad (5)$$

where $c = 2.99 \times 10^{10}$ cm s$^{-1}$ denotes the velocity of light.

Here, in the simulation, the pump excitation takes place from the first vibrational level ($v'' = 0$) of ground state X to the excited state B. The $\Delta t$ of the pump is considered as 50 fs throughout, which is shorter than the vibrational time period of oscillation in the B state (~300 fs), so that the excitation remains in the impulsive limit. The value of $E_0$ is taken as $1.23 \times 10^9 \, V/m$ (0.0024 in atomic unit).

## 2.2 Methods

### 2.2.1 WP dynamics

The pump-probe signal can be understood as the overlap of a time-dependent first-order 'bra' WP with a time-dependent second-order 'ket' WP [10,11]. The first order WP created by the first interaction of the pump pulse $E_{pump}(t)$ (at $t = t''$) can be written as,

$$|\psi^{(1)}(r,t)\rangle = \frac{i}{\hbar} \int_{-\infty}^{t} dt'' \left\{ e^{-\frac{iH_B(t-t'')}{\hbar}} [\mu_{BX} E_{pump}(t'')] e^{-\frac{iH_X t''}{\hbar}} \right\} |\psi^{(0)}(0)\rangle \quad (6)$$

Here, the phenomena after the first interaction of the pump pulse $E_{pump}(t)$ can be easily understood if this equation (6) is followed from the right-hand side. Initially, the ground state wavefunction $|\psi^{(0)}(0)\rangle$ (i.e. $v'' = 0$ of X state) evolves under the ground state Hamiltonian $H_X$, and the first interaction from the pump $E_{pump}(t'')$ creates the transition dipole moment ($\mu_{BX}$) coupling the ground state X with the first excited state B. As the newly created wavefunction evolves in the excited state Hamiltonian $H_B$, the integration takes care of all the instances of electric field interaction, which eventually gives the first-order excited state wave function $|\psi^{(1)}(r,t)\rangle$. The second-order WP created by the second interaction of the pump pulse $E_{pump}(t)$ (at $t = t''$) and the interaction of the probe pulse $E_{probe}(t)$ (at $t = t'$), can be written as,

$$|\psi^{(2)}(t)\rangle$$
$$= \left(\frac{i}{\hbar}\right)^2 \int_{-\infty}^{t} dt' \int_{-\infty}^{t'} dt'' \left\{ e^{-\frac{iH_E(t-t')}{\hbar}} [\mu_{EB} E_{probe}(t')] e^{-\frac{iH_B(t'-t'')}{\hbar}} [\mu_{BX} E_{pump}(t'')] e^{-iH_X t''/\hbar} \right\}$$
$$\times |\psi^{(0)}(0)\rangle \quad (7)$$

Similar to equation (6), equation (7) can also be understood following it from the right hand side. Initially, the ground state wavefunction $|\psi^{(0)}(0)\rangle$ (i.e. $v'' = 0$ of X state) evolves under the ground state Hamiltonian $H_X$, and the second interaction from the pump ($E_{pump}(t'')$) creates the transition dipole moment ($\mu_{BX}$) coupling the ground state X with the first excited state B. From $t = t''$ to $t = t'$, it evolves under the Hamiltonian of the B state $H_B$. At $t = t'$, the probe ($E_{probe}(t')$) creates the transition dipole moment ($\mu_{EB}$) coupling B with the second excited state E. Then, the newly created wavefunction evolves under the Hamiltonian of E state $H_E$ till time $t$. The double integral reflects the coherent contribution from all possible instances of the two field interactions at two different times. Therefore, the third-order polarization is the overlap of these first and second-order WPs, and can be written as,

$$P^{(3)}(t:t',t'') = \langle \psi^{(1)}(t)|\mu_{EB}^*|\psi^{(2)}(t)\rangle \tag{8}$$

The signal field, $E_{sig}(t:t',t'')$, is proportional to this third-order polarization, and is plotted in **Figure 2** for a given pump-probe delay, $T = t' - t'' = 400\ fs$ for a specific pump and probe pulse parameters as discussed later (in section 2.2.2.). In practice, the pump-probe signal is typically detected using heterodyne detection, where the signal field interferes with the transmitted probe field (which serves as a local oscillator for heterodyne detection). Therefore, the spectrum as a function of detection frequency (and pump-probe delay) is given by:

$$S(\omega_{det}, T) = |\tilde{E}_{det}(\omega_{det})|^2$$

$$= \left|FT\left(E_{sig}(t:t',t'') + E_{probe}(t')\right)\right|^2$$

$$= |\tilde{E}_{sig}(\omega_{det}, T) + \tilde{E}_{probe}(\omega_{det}, T)|^2 \tag{9}$$

Since the third-order signal field is too weak compared with the probe field, and in a pump-probe experiment, the probe-only signal is subtracted (say, by blocking every alternate pump pulse), the relevant term to be considered is:

$$(\omega_{det}, T) \approx \tilde{E}^*_{sig}(\omega_{det}, T) \times \tilde{E}_{probe}(\omega_{det}, T) + c.c.$$

$$= 2 \times Re\left(\tilde{E}^*_{sig}(\omega_{det}, T) \times \tilde{E}_{probe}(\omega_{det}, T)\right) \tag{10}$$

The spectrum, as a function of pump-probe delay, is plotted in **Figure 3**. The spectrally-integrated signal, detected by a point detector, can be obtained as:

$$S(T) = \int_0^\infty S(\omega_{det}; T) d\omega_{det} \tag{11}$$

This would be the same (barring a constant multiplicative factor) as the time-integrated signal:

$$S(T) \equiv I(T) = \int_{-\infty}^\infty 2 \times Re\left(E^*_{sig}(t:t',t'') \times E_{probe}(t')\right) dt \tag{12}$$

Both these signals, as a function of pump-probe delay, are plotted in **Figure 4**. For simplicity, we now analyze the signal I($T$) which contains both population information and coherence dynamics. The population dynamics is fitted with the population kinetics and subtracted from the signal to get the pure coherence dynamics.

Here, the pump-probe signal contains both the population as well as the vibrational coherence dynamics in B state, the population dynamics is subtracted after fitting the signal with the population kinetics using the following equation,

$$\frac{M}{2}\left(1 + erf\left[\frac{(t-0)}{\sigma_{IRF}}\right]\right), \tag{13}$$

The last equation is known as the cumulative distribution function that is obtained from the convolution two Gaussian functions (as pump and probe both have transform limited Gaussian profiles). Here, $M$ refers to the amplitude of the oscillation, and $\sigma_{IRF}$ corresponds to the standard deviation of convolution between pump and probe defined as,

$$\sigma_{IRF} = \sqrt{\frac{1}{2}\left[\left(\sigma_{pump}\right)^2 + \left(\sigma_{probe}\right)^2\right]}, \tag{14}$$

where $\sigma_{pump}$ and $\sigma_{probe}$ are the standard deviations of the field envelope of the pump and probe pulses, respectively.

### 2.2.2. ISRS

Let us consider an electric field $E(t)$ couples the ground vibronic state $|g\rangle|v''\rangle$ to the excited vibronic state $|e\rangle|v'\rangle$. Then, the population and coherence at time $t$ can be written as the following,

$$\int_{-\infty}^{t} |\langle v'|\langle e|\mu E(t')|g\rangle|v''\rangle|^2 dt' \tag{15}$$

Following an impulsive excitation, the Franck-Condon overlap between the vibrational levels is independent of $t$ and can be taken out of the integration. Then, equation (15) becomes,

$$|\langle v'|v''\rangle|^2 \int_{-\infty}^{t} |\langle e|\mu E(t')|g\rangle|^2 dt' \tag{16}$$

$$= |\mu_{eg}|^2 |\langle v'|v''\rangle|^2 \int_{-\infty}^{t} |E(t')|^2 dt' \tag{17}$$

where $\mu_{eg} = \langle e|\mu|g\rangle$ is the electronic transition dipole moment. Now, from equation (17), the excited state population is,

$$\sum_{v\prime} |\mu_{eg}|^2 \left[\int_{-\infty}^{t}|E(t')|^2 dt'\right] |\langle v'|v''\rangle|^2 \tag{18}$$

Now, The coherence between two vibrational levels $|v_j'\rangle$ and $|v_k'\rangle$ in the excited state (as shown in **Figure 5**), from equation (18), can be obtained as,

$$A = |\mu_{eg}|^2 \left[\int_{-\infty}^{t}|E(t')|^2 dt'\right] \left\{\langle v''|v_j'\rangle\langle v_k'|v''\rangle\, e^{-i\left[\frac{(E_j - E_k)t}{\hbar}\right]} + \text{c. c.}\right\} \tag{19}$$

$$= 2|\mu_{eg}|^2 \left[\int_{-\infty}^{t}|E(t')|^2 dt'\right] \left\{\langle v''|v_j'\rangle\langle v_k'|v''\rangle \cos\left[\frac{(E_j - E_k)t}{\hbar}\right]\right\} \tag{20}$$

Considering all the possible combinations in the excited state vibrational levels and the spectral amplitude of the pump interaction $E_{pump}(t)$, from equation (20), the overall vibrational coherence dynamics for a transition from ground state X to excited state B can be written as,

$$C = \sum_{j \neq k} 2|\mu_{BX}|^2 \tilde{E}_{pump}(\omega_j - \omega_{pump}) \tilde{E}_{pump}(\omega_k - \omega_{pump}) \left[\int_{-\infty}^{t} |E_{pump}(t')|^2 dt'\right] \left\{\langle v''|v_j'\rangle\langle v_k'|v''\rangle \cos\left[\frac{(E_j - E_k)t}{\hbar}\right]\right\}, \tag{21}$$

where $\tilde{E}_{pump}(\omega_j - \omega_{pump})$ denotes the spectral field amplitude of pump at $\omega_j$.

Therefore, for a transition from ground state X to two levels $|v_j'\rangle$ and $|v_k'\rangle$ in excited state B, the last expression is written as,

$$C_{jk} = 2|\mu_{BX}|^2 \tilde{E}_{pump}(\omega_j - \omega_{pump}) \tilde{E}_{pump}(\omega_k - \omega_{pump}) \left[\int_{-\infty}^{t} |E_{pump}(t')|^2 dt'\right] \left\{\langle v''|v_j'\rangle\langle v_k'|v''\rangle \cos\left[\frac{(E_j - E_k)t}{\hbar}\right]\right\} \tag{22}$$

For a pump at 587 nm (17,035 cm$^{-1}$) with 50 fs intensity FWHM (416 cm$^{-1}$ FWHM for $\tilde{E}(\omega)$), the vibrational coherences are created covering around 7 vibrational levels ($v' = 8$ to 14) in the B state that oscillate with time. Now, for an excited-state absorption process, these coherences are projected to a higher-lying electronic state (E) following an interaction of the probe pulse. The range for the probe frequencies is chosen by taking the maximum energy gaps between the higher-lying excited state (E) and the highest ($v' = 14$) and the lowest ($v' = 8$) vibrational levels that are involved in the pump-induced coherences in the B state, and the maximum ($\omega_1$) and minimum ($\omega_2$) probe frequencies are calculated from the inner turning points of the lower level ($v' = 8$), and the outer turning point of the higher level ($v' = 14$) in the B state, respectively, as shown in **Figure 6a**. This choice corresponds to a probe pulse of 2 fs duration, which is unrealistic from the experimental point of view. For this reason, the maximum and minimum energy gaps are chosen from the outer turning points of the corresponding vibrational levels, as shown in **Figure 6b** (following reference 48 [48]). Thus, we project a part of the vibrational WP to 'E' state by using a longer probe pulse of ~18 fs duration (1,156 cm$^{-1}$ FWHM of $\tilde{E}(\omega)$) and centred on 406 nm (24,630 cm$^{-1}$).

In general, the pump-probe spectra are obtained by time-delaying the probe. However, in our method, each of the coherences generated due to the pump excitation, as represented by $C_{jk}$, is a time-dependent function and therefore we don't need to explicitly include the pump-probe delay. These coherences are now projected to the 'E' state through the corresponding FC factor between the vibrational level $|v_j'\rangle$ and $|v_e\rangle$ of the respective first (B) and second (E) excited states, the transition dipole moment between then, and the spectral amplitude of the probe pulse $E_{probe}(t)$ as:

$$C_{jk}|\langle v_e|v_j'\rangle|^2 \mu_{EB} \tilde{E}_{probe}(\omega_{ej} - \omega_{probe}) \quad (23)$$

where $\tilde{E}_{probe}(\omega_{ej} - \omega_{probe})$ denotes the spectral field amplitude of probe at $\omega_{ej}$, the energy difference between $|v_e\rangle$ and $|v_j'\rangle$.

As the pump-probe signal is self-heterodyned with the probe, the overall signal for a particular set of coherence between $|v_j'\rangle$ and $|v_k'\rangle$ can be obtained as:

$$C_{jk}|\langle v_e|v_j'\rangle|^2 \tilde{E}_{probe}(\omega_{ej} - \omega_{probe})|\langle v_e|v_k'\rangle|^2|\mu_{EB}|^2 \tilde{E}_{probe}(\omega_{ek} - \omega_{probe}) \quad (24)$$

The last equation explains either the probe transition starts from $|v_j'\rangle$ and the signal ends in $|v_k'\rangle$, giving rise to the Stokes pathway of pump-probe signal, as shown in **Figure 5a**. Alternatively, where the probe excitation happens from $|v_k'\rangle$, then the corresponding signal ends in $|v_j'\rangle$ is represented as the coherent anti-Stokes pathway, as shown in **Figure 5b**. It is important to note that even if the amplitudes of both the Stokes and coherent anti-Stokes pathway are represented by the last equation, there is an inherent $\pi$ phase-shift between these two pathways. The reason being the interaction of probe pulse from the ket (or bra) side of energy level diagram for the Stokes (or coherent anti-Stokes) pathway. Therefore, if the last equation represents the signal for Stokes pathway, then the signal for the coherent anti-Stokes pathway can be written as:

$$C_{jk}|\langle v_e|v_k'\rangle|^2 \tilde{E}_{probe}(\omega_{ek} - \omega_{probe})|\langle v_e|v_j'\rangle|^2 |\mu_{EB}|^2 \tilde{E}_{probe}(\omega_{ej} - \omega_{probe})e^{i\pi} \quad (25)$$

Now, for a set of $j$ and $k$ values, the Stokes and coherent anti-Stokes signal appear at different detection wavelengths (i.e. $\omega_{ek}$ for the Stokes or $\omega_{ej}$ for the coherent anti-Stokes). For all combinations of $j$ and $k$ values, all these pathways together constitute the pump-probe spectra.

### *2.2.3. Simulation methods*

Here, we numerically perform the WP propagation using the split-operator method with a sampling rate of $0.1 fs$ to obtain the time-dependent wave functions in both the the excited-states. The coordinate space is chosen as a grid of -10 Å to 10 Å with 512 grid points, is defined in a column vector. The entire simulation study and data plotting have been carried out using MATLAB programming (version 2023a, MathWorks Inc.)

## 3. Results and discussion

### *3.1 WP dynamics*

The pump-probe signal is first simulated for the interference of the first- and second-order WP approach with 50 fs pump and 18 fs probe excitation (corresponding to $\sigma_{pump} = 30$ fs and $\sigma_{probe} = 11.3$ fs, in equation (14)) for 2 ps pump-probe delay. **Figure 7a** shows that the signal starts from zero and then oscillates periodically with an oscillation period of around 300 fs, similar to the vibrational time-period of the B state. However, there is significant modulation in the signal that signifies the involvement of multiple vibrational levels in the B state due to pump-pulse excitation.

After subtracting the population kinetics using equation (13), the residual signal that only contains the coherence dynamics is shown in **Figure 7b** and is used to analyse the excited-state dynamics in further discussions.

Note that, in this method, the signal is collected as a function of pump-probe delay at only a fixed probe wavelength. Therefore, this method is not enough to observe the signal in a way where all the probe wavelengths are frequency dispersed to distinguish the Stokes and coherent anti-Stokes signal. For that we move to the second method that follows an analytical approach considering all individual vibrational levels in the B state that contain multiple vibrational coherences among themselves, upon pump excitation.

*3.2 ISRS*

First, we consider the case where the coherences are created between two adjacent vibrational levels (i.e. $\Delta v = \pm 1$) under the spectral envelope of the pump excitation. For a pump excitation of 50 fs FWHM at 587 nm, all the $\Delta v = \pm 1$ coherences created between the lower vibrational level $v' = 8$ and higher vibrational level $v' = 14$ (e.g. coherences between $v' = 8$ and 9, $v' = 9$ and 10 etc.) are projected to the higher lying electronic state 'E' by the probe pulse, and after addition of all these contributions, the signal for the ESA pathway is obtained as a function of the pump-probe delay as well as detection wavelength of the probe in a frequency dispersive manner. It is observed from **Figure 8** that the Stokes and the coherent anti-Stokes signals appear at different regions of the detection axis, which directly measures the frequency offset between these two signal pathways. Now after the addition of these two signal pathways, the overall pump-probe spectrum is obtained, as shown in **Figure 8c**. Also, a nodal structure is observed at around 406 nm, corresponding to the central frequency of the probe pulse, due to the anti-phased oscillation in the stokes and coherent anti-Stokes signals. Thus, these two signal pathways are isolated at either region around the node.

Now, it is important to note that the magnitude of the signals from these two pathways are exactly equal, however, differ being either positive or negative. Therefore, in **Figure 8c**, if the projection along the detection wavelength is taken, the overall spectral intensity vanishes as they nullify each other's contribution. To check how this second methods resembles with the first method, the projection is taken along the detection wavelength for the signal contribution coming from only the coherent anti-Stokes pathway to get the signal only as the function of pump-probe delay. This is now compared to the result obtained from the WP dynamics approach. **Figure 9** shows that for the $\Delta v = \pm 1$ coherences, it only shows periodic oscillations without any modulation as observed in **Figure 7b**. This hints or emphasizes the need for considering contributions from other type of coherences such as $\Delta v = \pm 2$ or $\Delta v = \pm 3$ in addition to the fundamental one.

**Figure 10** shows the similar pump-probe spectra for $\Delta v = \pm 2$ coherences, where the signal oscillates periodically, however, not as clearer as it is observed for the case of $\Delta v = \pm 1$. It is also observed that the intensity of the signals is quite less as compared to the previous case. Again, similar as earlier, a projection along the detection wavelength for the coherent anti-Stokes pathway is taken and the one-dimensional result (as shown in **Figure 11**) is compared to the result obtained from the WP dynamics approach.

Now, it is observed that there are modulation in the signal obtained from the analytical method after considering both $\Delta v = \pm 1$ and $\Delta v = \pm 2$ coherences, as observed in **Figure 7b** for the WP dynamics approach.

For a better comparison of the results as obtained from both the methods, the contributions due to $\Delta v = \pm 3$ coherences are also simulated using the second method, as shown in **Figure 12**, and when added to the results of the other two types of coherences as found earlier, in **Figure 13**, the second method shows more agreement with the results as obtained in the WP dynamics method.

## 4. Further discussion

It is important to note that the results of these two methods, as discussed in the last section, still does not match exactly. This is likely due to the consideration of coherences in B state lying only within the FWHM of the pump pulse, which we now discuss in the following.

In section 3.1, the pump-probe signal is simulated till 2 ps delay where the pure vibrational coherence dynamics shows modulated oscillatory feature explaining the involvement of different types of coherences i.e. $\Delta v = \pm 1$, $\Delta v = \pm 2$, and $\Delta v = \pm 3$ coherences etc. Now, to get a clear idea, the pump-probe spectra is further simulated for longer pump-probe delay of 20 ps following the WP dynamics approach. It is observed in **Figure 14a** that at around 18 ps, the coherent oscillation revives back, confirming the half revival time for B state of iodine. Following the similar approach as mentioned in section 3.1, the pure vibrational coherences are extracted after fitting and subtracting the population kinetics from the pump-probe spectra as shown in **Figure 14b**. Upon Fourier transforming the temporal oscillation, the vibrational modes are obtained as shown in **Figure 14c** that clearly shows the contributions coming from $\Delta v = \pm 1, \Delta v = \pm 2, \Delta v = \pm 3$ and $\Delta v = \pm 4$ coherences containing vibrational substructures with varying amplitudes. To further confirm the vibrational modes, the temporal signal is processed by using the Hanning window that is used to remove the side bands and it smoothly damps the oscillation at 20 ps, as shown in **Figure 15a**. The Fourier transformed data of this processed signal is shown in **Figure 15b** and it is observed that the maximum contribution towards the vibrational coherence is arising from the vibrational mode at around 106 cm$^{-1}$ of the $\Delta v = \pm 1$ coherences. This particular mode is equivalent to the creation of vibrational WP between the vibrational levels $v'_j =12$, and $v'_k =13$ at respective potential energies of 17,120 cm$^{-1}$ and 17226 cm$^{-1}$ in the B state, as denoted by $E_{v'_j}$ and $E_{v'_k}$ in **Figure 16**.

Now, to ascertain whether the signal is originating due to the Stokes or coherent anti-Stokes pathway, a qualitative analysis is performed. First, the potential energy values of the E state are noted at the coordinate specified by the outer turning points of both the vibrational levels $v'_j =12$ and $v'_k =13$ in the B state. The potential energies are obtained as 41,860 cm$^{-1}$ at $r_1 = 3.443$ and 41,754 cm$^{-1}$ at $r_2 = 3.467$ from the E state PES, as denoted by $E_{e_j}$ and $E_{e_k}$ in **Figure 16**. As mentioned earlier that probe wavelength is also chosen from the outer turning points of the B state, next, the equivalent probe energy of 24,618 cm$^{-1}$ for the central wavelength of 406.2 nm is added vertically to the individual potential energies of the vibrational levels. Thus, after addition of the probe energy, the overall potential energies becomes 41,738 cm$^{-1}$ for the transition from $v'_j =12$ and 41,844 cm$^{-1}$ for the transition from $v'_k =13$. After comparing these energy values with $E_{e_j}$ and $E_{e_k}$, it is clear that the probability of the probe transition from the outer turning point of $v'_k =13$ level is larger (shown with darker green arrow for probe transition in Figure 13) than the other vibrational level ($v'_j =12$) as it surpasses the potential energy in the E state. Therefore, as the probe transition is more favourable from the upper vibrational level in B state, the pump-probe data obtained from WP dynamics approach is majorly contributed by the coherent anti-Stokes signal.

## 5. Conclusion

In summary, we have shown how coherent excited-state vibrational dynamics simulated through interference between first- and second-order WPs can be correlated to Stokes and coherent anti-Stokes pathways using a state-to-state excitation framework. Based on this simplified model system, like molecular I$_2$, a similar study can be further extended to complex molecular systems, such as molecules having double minima structures in their excited electronic states, showing contemporary applications in quantum information science [49], under a pulse-pair excitation approach, known as linear WP interferometry [50].

**Figures**

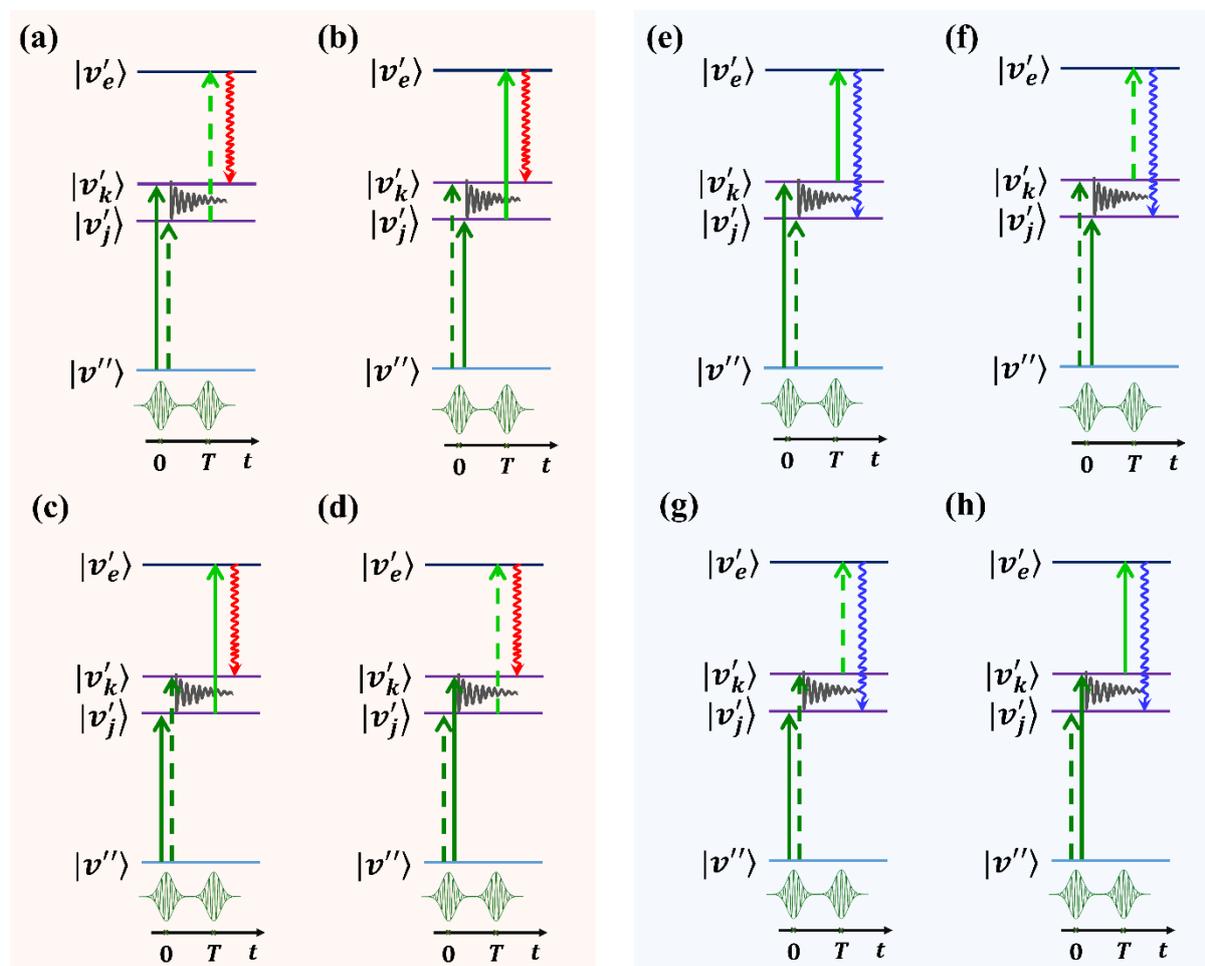

**Figure 1.** (a to h) Energy level diagrams showing coherence dynamics under resonant pump-probe excitation for excited state absorption process. First a two-field excitation (at $t = 0$) creates vibrational coherence between two vibrational levels in the excited state that is further probed to higher lying excited state as a function of pump-probe delay $T$. The (vertical) solid/broken arrow corresponds to interaction of the field with bra/ket side of the density matrix while (vertical) wavy arrow (red/blue) corresponds to the signal. Time propagates (shown with black arrows) from left to right for each of the energy level diagrams (cross marker dictates the time of electric field interaction). Left panel (in red) corresponds to the Stokes pathway, and right panel (in blue) corresponds to the coherent anti-Stokes pathway. Note that all the pathways are shown with its complex conjugate side-by-side (i.e. the complex conjugate of 'a' or 'c' are shown in 'b' and 'd', respectively).

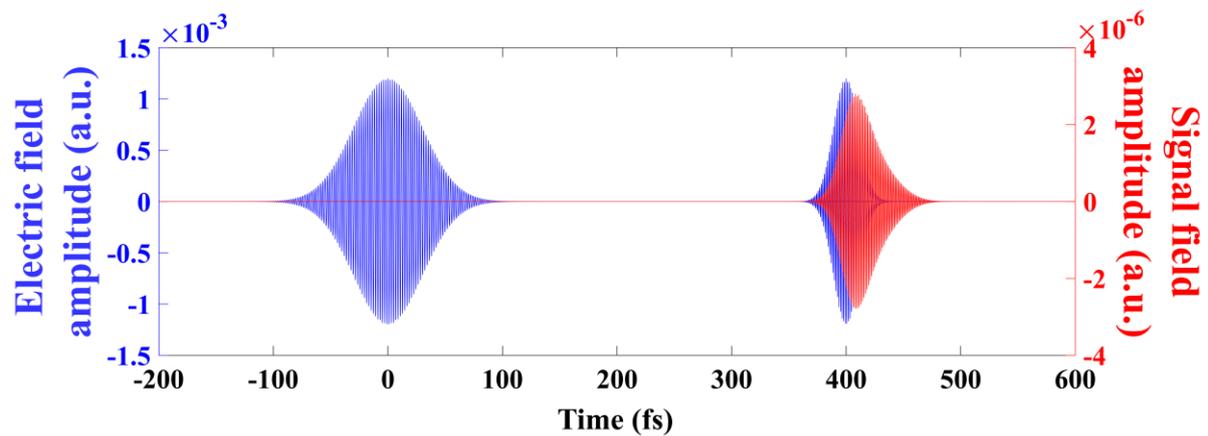

**Figure 2.** Time domain representation of a 50 fs pump, 18 fs probe (in blue), and the signal (in red) equivalent to the third order nonlinear polarization generated at 400 fs pump-probe delay.

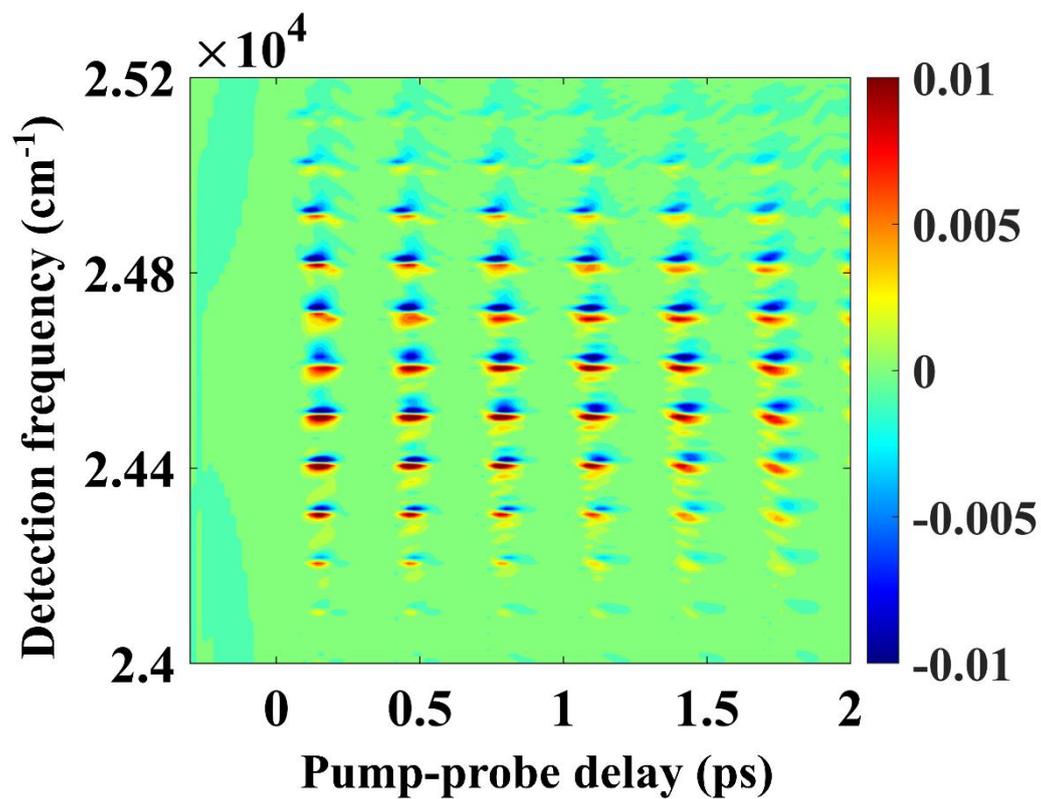

**Figure 3.** Pump-probe contour map as a function of detection frequency (cm$^{-1}$) and pump-probe delay (ps) for a 50 fs pump and 18 fs probe excitation.

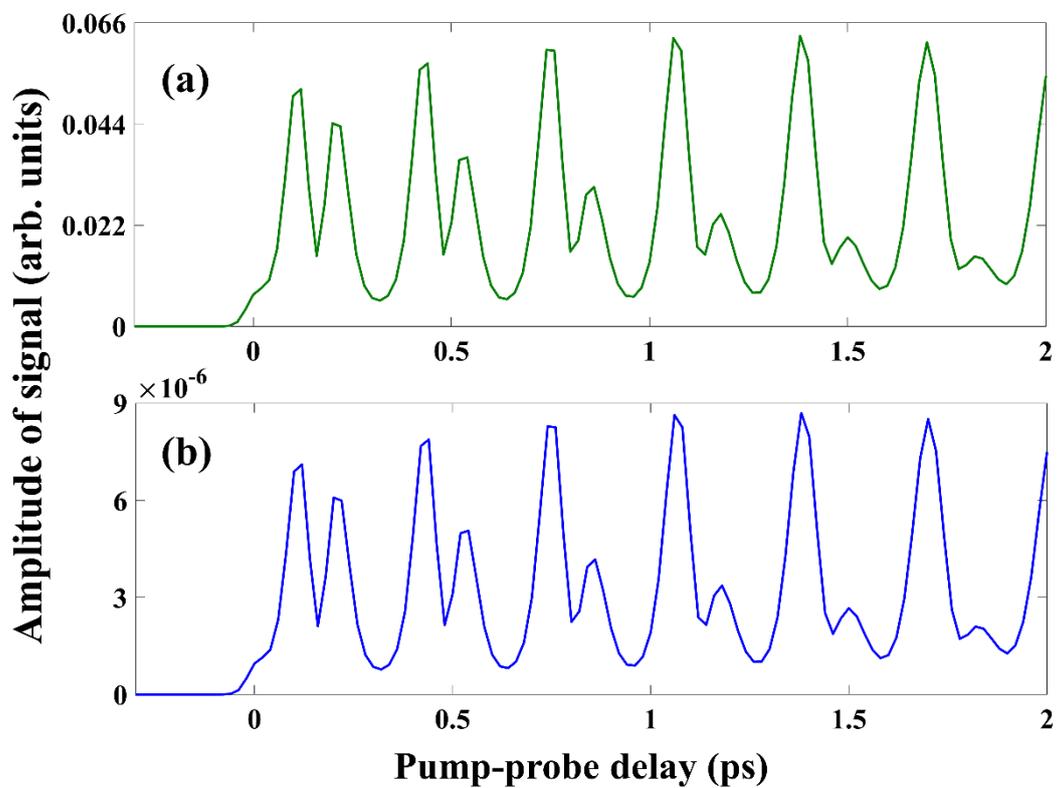

**Figure 4.** (a) Spectrally integrated signal along the detection frequency of the pump-probe spectrum. (b) Time-integrated pump-probe signal as measured by a point detector. Both of the signals are obtained for a 50 fs pump and 18 fs probe excitation, and represent the same dynamics differing by a constant multiplicative factor.

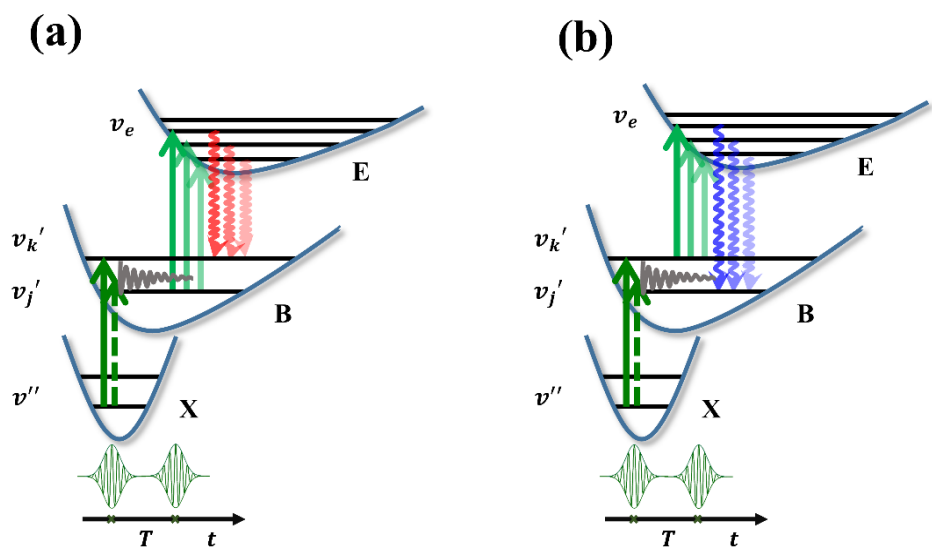

**Figure 5.** Schematic representation of creation of vibrational coherences (grey oscillation) in the B state following two-field interaction (solid/broken arrow in green) from the pump. After a time-delay probe pulse projects these coherences to E state and the signal is emitted either following (a) Stokes or (b) coherent anti-Stokes pathway.

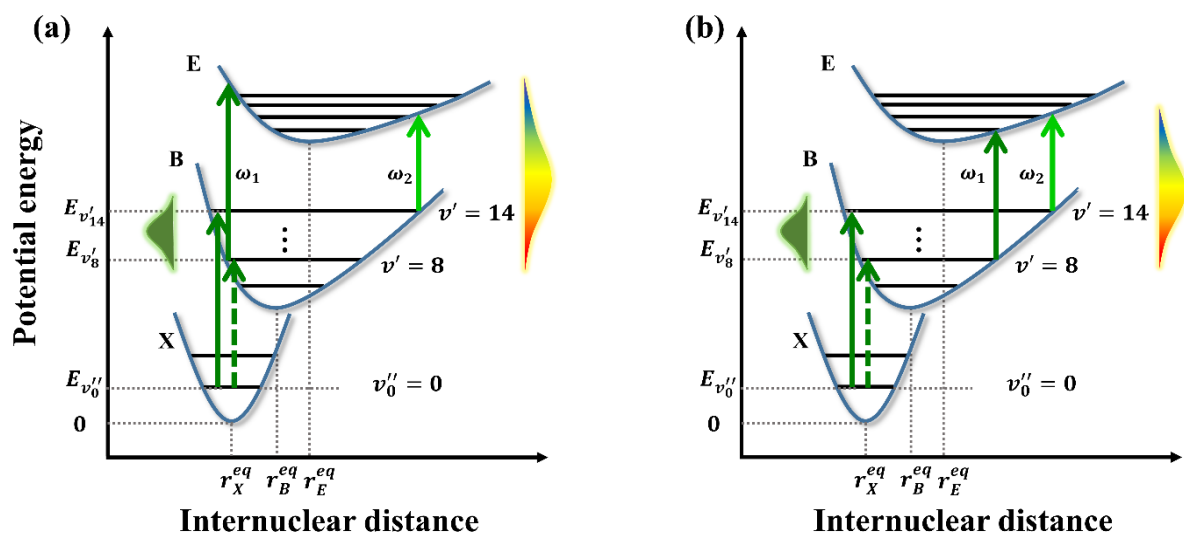

**Figure 6.** Schematic illustration of probe wavelength selection. The maximum and minimum vibrational levels in the B state that contribute to vibrational coherence, created via pump excitation from the X state (solid/broken arrows), are indicated by their respective energies. (a) The probe energies, $\omega_1$ and $\omega_2$, correspond to transitions from the inner turning point of the lower vibrational level and the outer turning point of the upper vibrational level in the B state, respectively. (b) The probe energies, $\omega_1$ and $\omega_2$, correspond to transitions from the inner turning points of both the lower and upper vibrational levels in the B state. Note that the PESs and vibrational levels are shown schematically and are not to scale with actual molecular parameters.

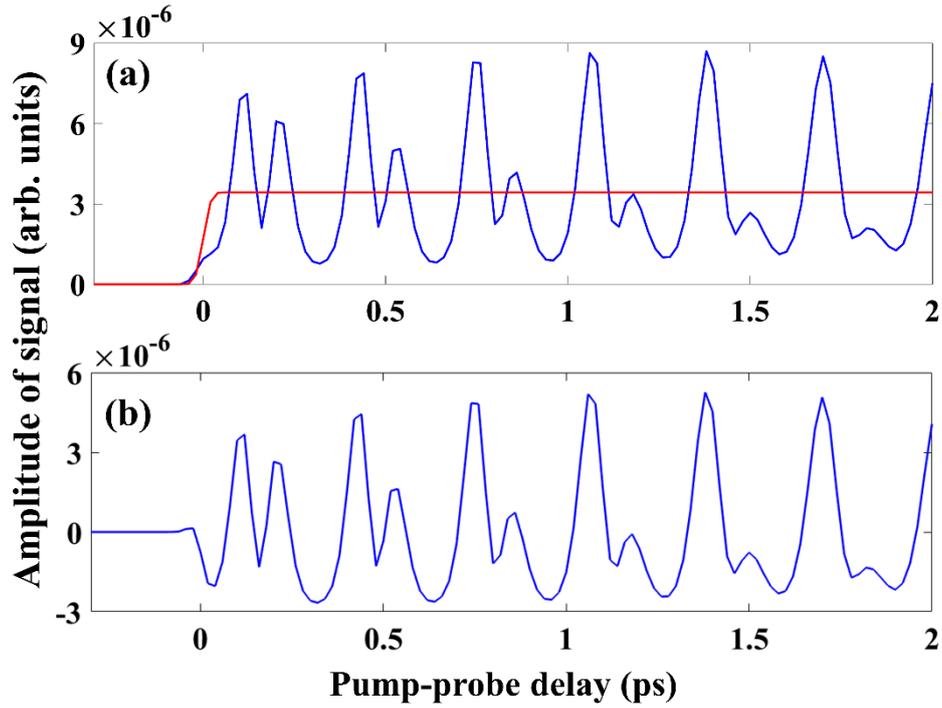

**Figure 7.** (a) Pump-probe signal (in blue) following WP dynamics method that contains both population and coherence dynamics. The fitted population kinetics is shown in red. (b) Pure vibrational coherence dynamics after subtracting the population kinetics from the pump-probe signal.

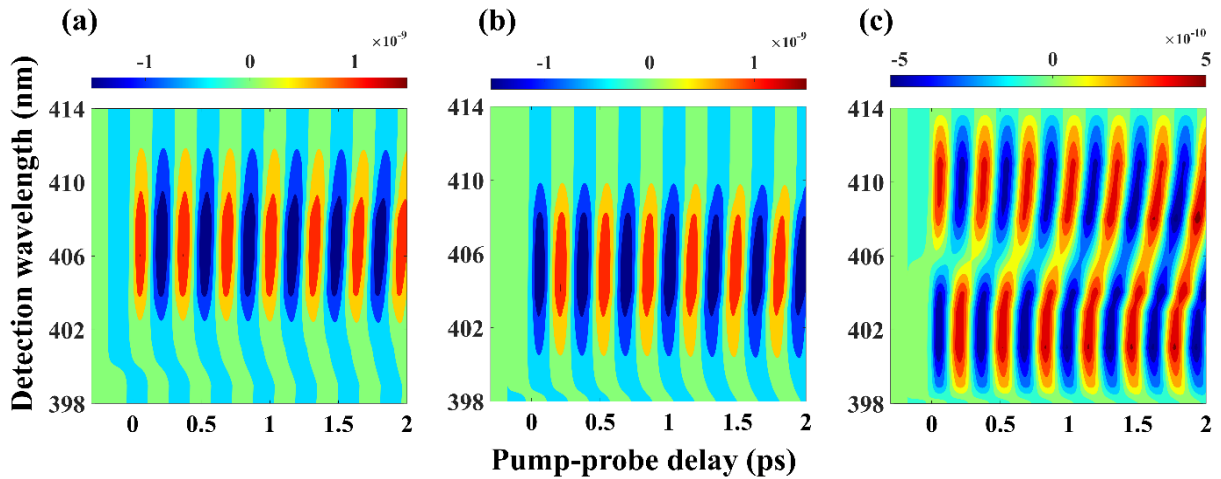

**Figure 8.** Pump-probe spectra for (a) Stokes, (b) coherent anti-Stokes pathways showing their appearances at different detection wavelengths considering only the $\Delta v = \pm 1$ coherences. (c) Addition of these two pathways represent the overall spectra that clearly shows the anti-phase relationship between (a) and (b).

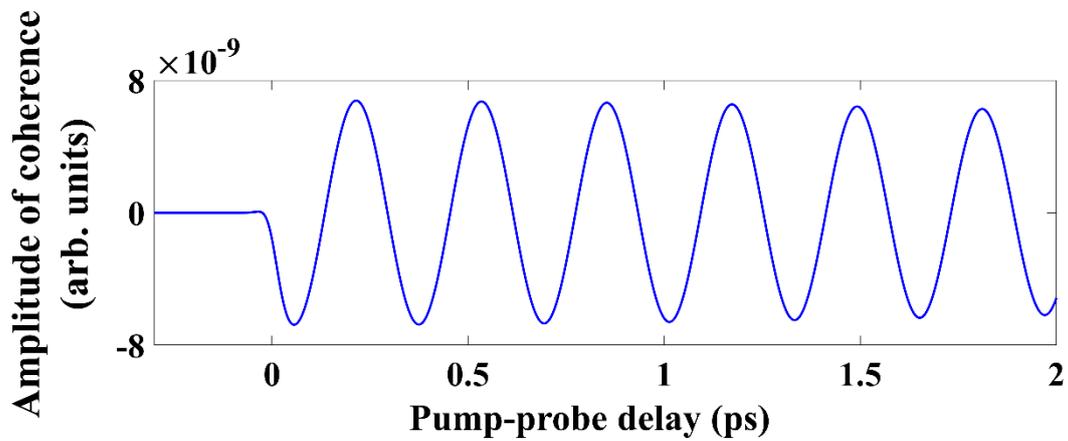

**Figure 9.** Projection along the detection axis of the coherent anti-Stokes signal containing only $\Delta v = \pm 1$ coherences.

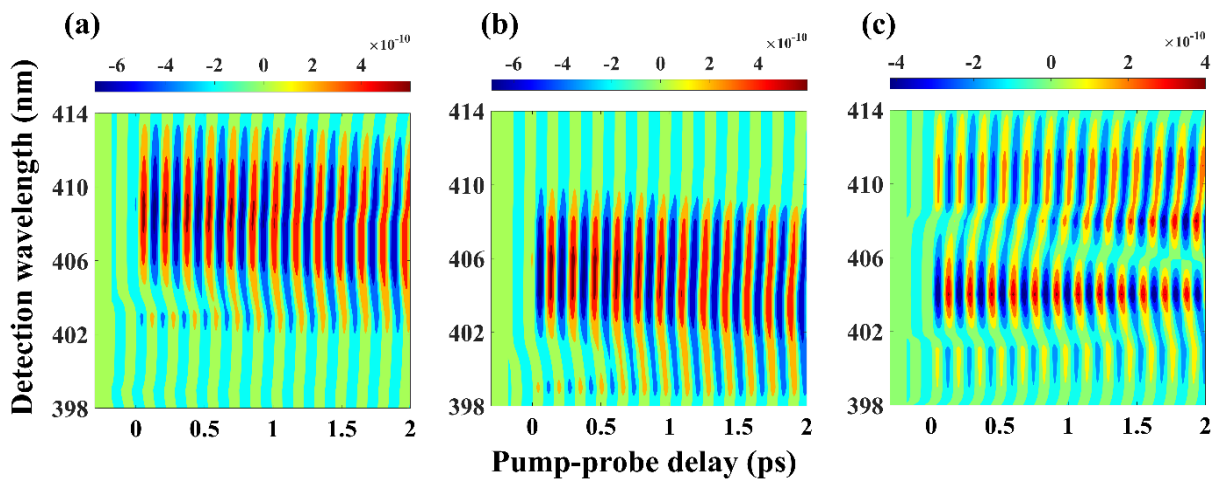

**Figure 10.** Pump-probe spectra for (a) Stokes, (b) coherent anti-Stokes pathways showing their appearances at different detection wavelengths considering only the $\Delta v = \pm 2$ coherences. (c) Addition of these two pathways represent the overall spectra that clearly shows the anti-phase relationship between (a) and (b).

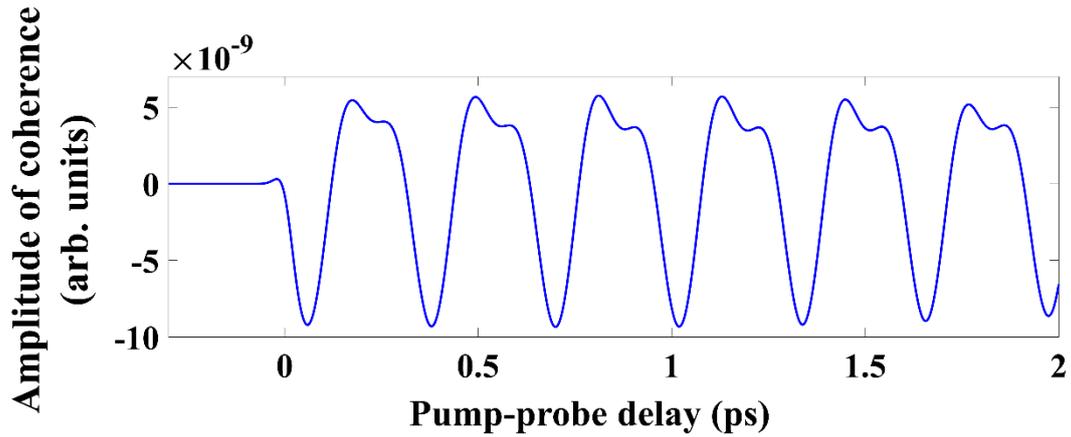

**Figure 11.** Projection along the detection axis of the coherent anti-Stokes signal containing $\Delta v = \pm 1$ and $\Delta v = \pm 2$ coherences.

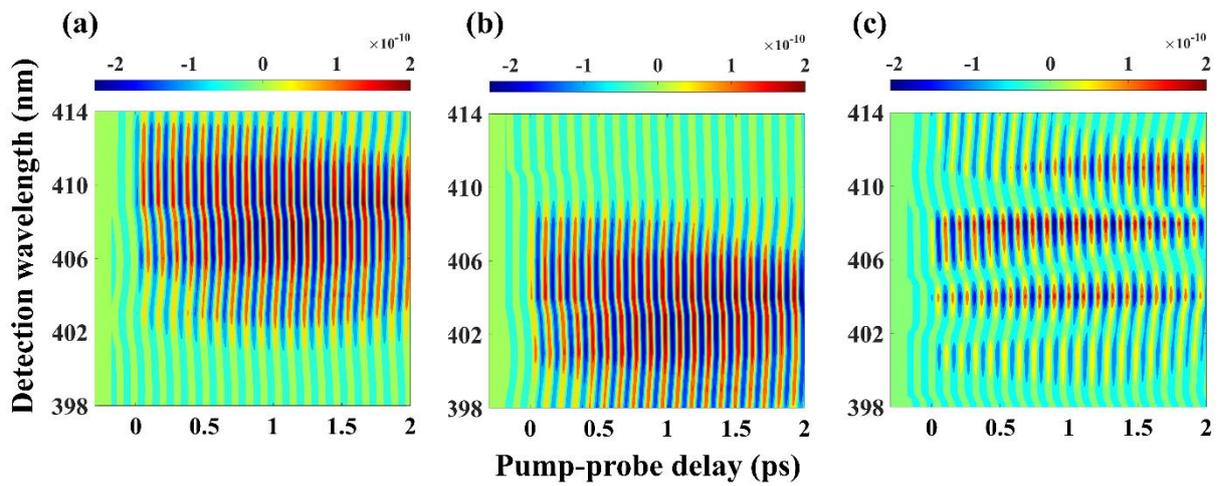

**Figure 12.** Pump-probe spectra for (a) Stokes, (b) coherent anti-Stokes pathways showing their appearances at different detection wavelengths considering only the $\Delta v = \pm 3$ coherences. (c) Addition of these two pathways represent the overall spectra that clearly shows the anti-phase relationship between (a) and (b).

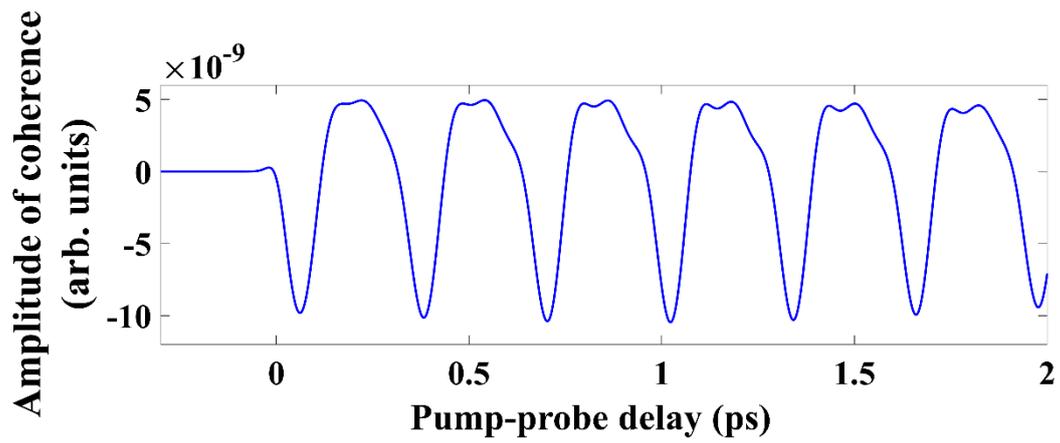

**Figure 13.** Projection along the detection axis of the coherent anti-Stokes signal containing $\Delta v = \pm 1$, $\Delta v = \pm 2$, and $\Delta v = \pm 3$ coherences.

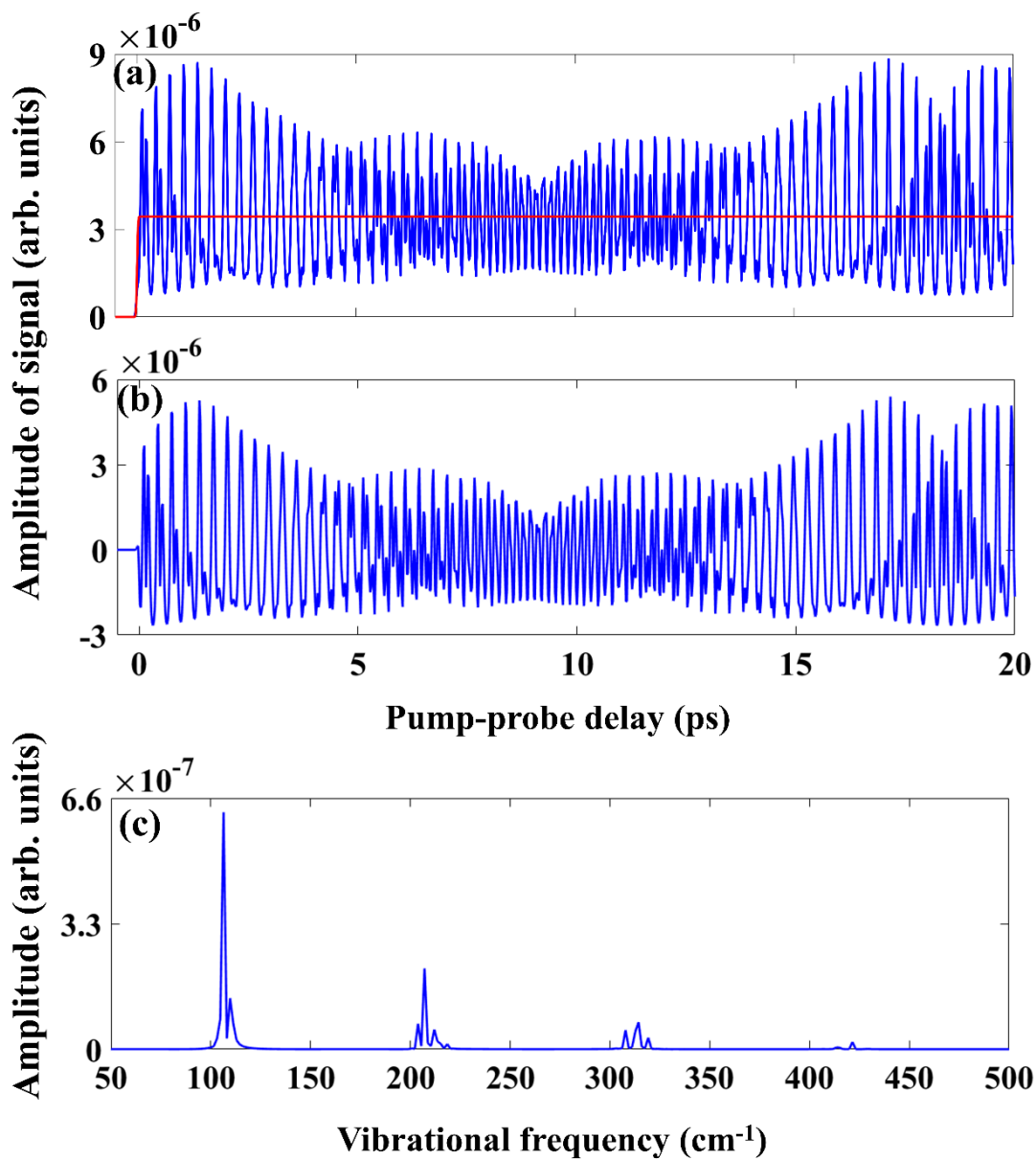

**Figure 14.** (a) Pump-probe signal (in blue) following WP dynamics method that contains both population and coherence dynamics. The fitted population kinetics is shown in red. (b) Pure vibrational coherence dynamics after subtracting the population kinetics from the pump-probe signal. (c) Vibrational modes that are obtained upon Fourier transforming the time-domain oscillation as shown in 'b'.

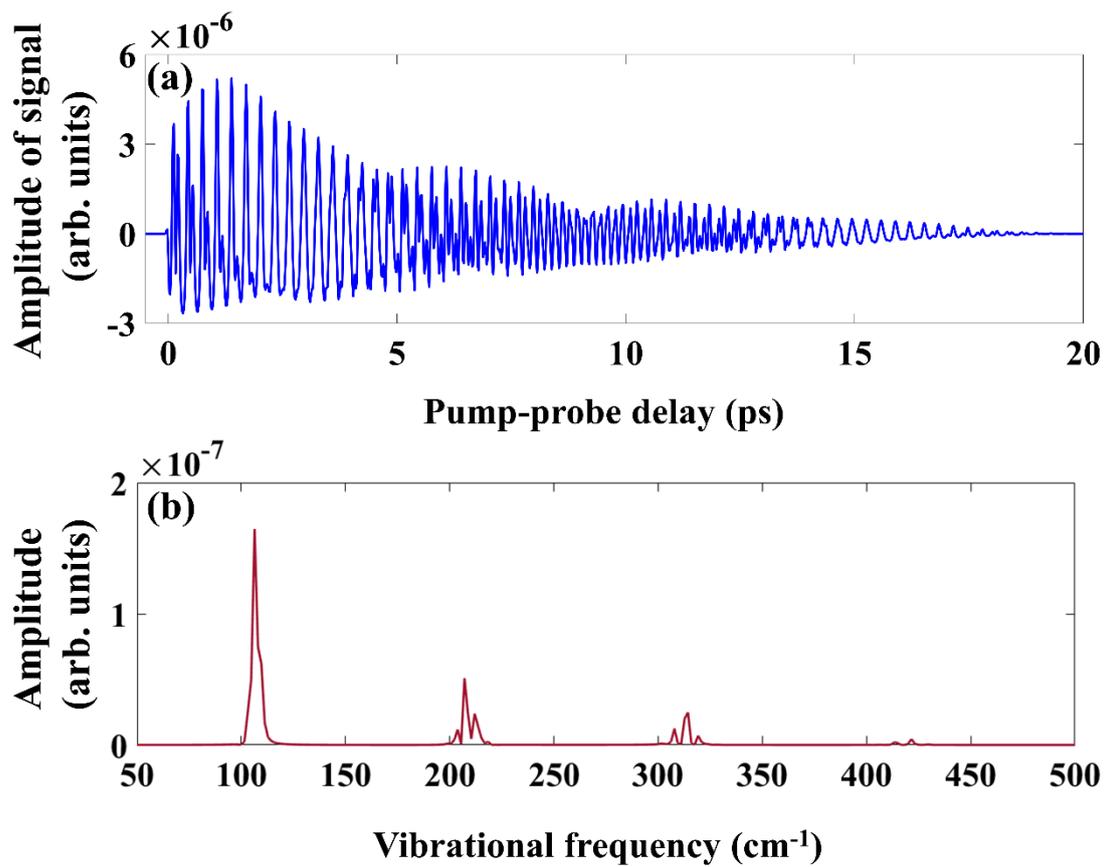

**Figure 15.** (a) Processes time-domain vibrational coherence dynamics and (b) corresponding vibrational modes after using Hanning window function in the raw data as shown in Figure 11b.

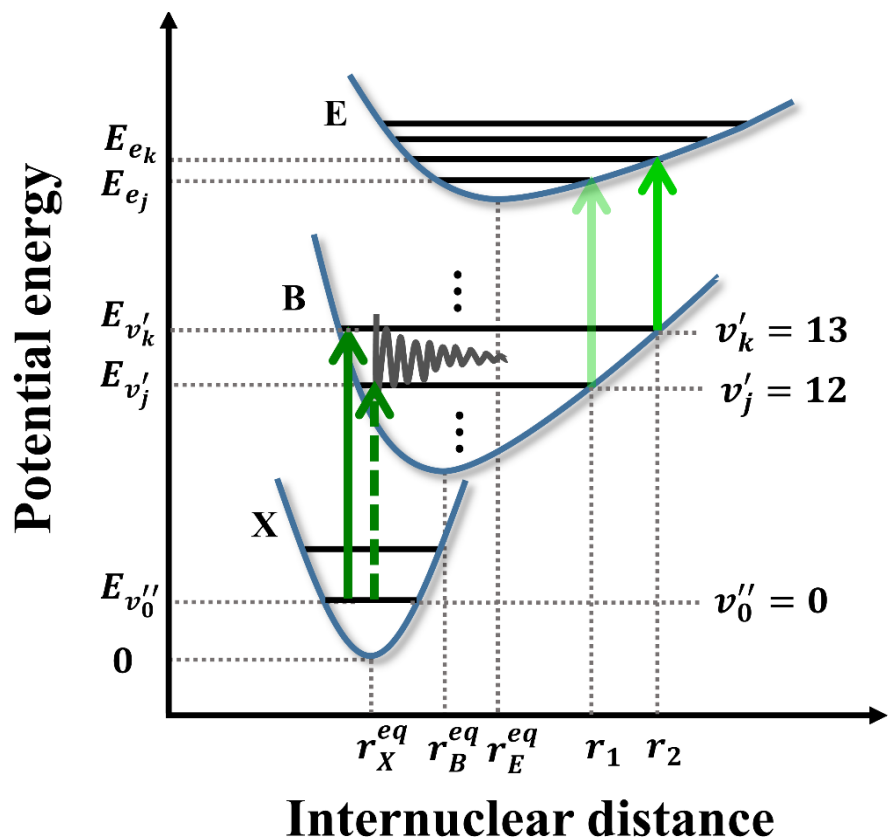

**Figure 16.** Schematic showing the major contribution towards the vibrational coherences in B state as obtained from WP dynamics method. The involved vibrational levels are denoted with specified energies in the vertical axis and corresponding nuclear coordinates are specified in the horizontal axis using the dotted lines. Note that the PESs and the vibrational levels are not positioned exactly as the original molecular parameters in this schematic.